\documentclass[11pt]{article}
\usepackage{aaspp4,flushrt,rotating,epsfig}

\def\etal{{\it et al.\ }}
\def\a3b{(\alpha+3\beta)}
\def\erf{\mathrm{erf}}
\def\be{\begin{equation}}
\def\ee{\end{equation}}
\def\mj{M_\mathrm{J}}
\def\half{{\textstyle{1\over2}}}
\def\ffrac#1#2{{\textstyle\frac{#1}{#2}}}
\def\xmax{x_\mathrm{max}}
\def\ymin{y_\mathrm{min}}
\def\pmax{P_\mathrm{max}}

\begin{document}

\title{Maximum-likelihood method for estimating the mass and period
distributions of extra-solar planets}

\author{Serge Tabachnik and Scott Tremaine}
\affil{Princeton University Observatory, Peyton Hall, Princeton, NJ
08544-1001, USA} 

\begin{abstract}

\noindent
We investigate the distribution of mass $M$ and orbital period $P$ of
extra-solar planets, taking account of selection effects due to the
limited velocity precision and duration of existing surveys. We fit
the data on 63 planets to a power-law distribution of the form
$dn=CM^{-\alpha}P^{-\beta}(dM/M)(dP/P)$, and find
$\alpha=0.12\pm0.10$, $\beta=-0.26\pm0.06$ for $M\lesssim 10\mj$,
where $\mj$ is the Jupiter mass. The correlation coefficient between
these two exponents is $-0.32$, indicating that uncertainties in the
two distributions are coupled. We estimate that 3\% of solar-type
stars have companions in the range $1\mj<M<10\mj$, $2\hbox{
d}<P<10\hbox{ yr}$.

\end{abstract}


\section{Introduction}
\noindent
As of May 2001, radial-velocity surveys have discovered over sixty
planets orbiting nearby stars. This sample should be large enough to
provide reliable estimates of the distribution of planetary mass and
orbital elements, at least in the range to which the radial-velocity
surveys are sensitive.

At least two important selection effects must be included in any
statistical analysis of this kind: (i) each survey has a detection
limit $K_D$, such that the orbits of companions that induce reflex
motions in their host star of amplitude $<K_D$ cannot be reliably
characterized; (ii) orbits of companions with periods much longer than
the duration of the survey cannot be reliably characterized. In any
survey limited by its velocity precision, uncertainties in the
distribution of planetary masses $M$ are coupled to uncertainties in
the distribution of orbital periods $P$, because the velocity
amplitude induced by a companion is $\propto MP^{-1/3}$
(eq.~\ref{eq:kdef}). Thus it is necessary to determine both
distributions simultaneously.

The aim of this paper is to describe a maximum-likelihood method of
estimating these distributions using data from a variety of surveys,
while accounting for survey-dependent selection effects.  We have
chosen to fit the data to simple power-law models of the distribution
of masses and periods; such distributions are simple to interpret and
common in nature, and it is straightforward to generalize our approach
to non-parametric models as the data improve.

We shall work with data from eight radial-velocity surveys of nearby
stars, which have detected between 2 and 23 extra-solar planets each
(see Table \ref{tab:surveys}). Together these surveys have detected 90
planets as of June 1 2001, although several planets appear in more
than one survey so we have only 63 distinct planets.

We compare our method and results with other determinations of the
mass distribution of extra-solar planets in \S\ref{sec:disc}.

\section{Maximum Likelihood Method}

\noindent
We focus initially on the simple case of a single survey that examines
$N^\ast$ stars for radial velocity variations due to orbiting
companions. The velocity amplitude $K$ due to a companion of mass $M$
with orbital period $P$ is
\be 
K={M\sin i\over M_\ast+M}(1-e^2)^{-1/2}\left[2\pi G(M_\ast+M)\over
P\right]^{1/3},
\label{eq:K}
\ee 
where $e$ is the orbital eccentricity, $M_\star$ is the stellar mass,
and $i$ is the inclination between the orbit plane and the sky
plane. For simplicity we assume that $e\ll1$ so that the factor
$(1-e^2)^{-1/2}$ is unity. We do not attempt to account more
accurately for the eccentricity dependence because the detectability
limit for eccentric orbits depends on both the amplitude and shape of
the radial-velocity curve. At the upper quartile of the eccentricities
in our sample, $e=0.46$, the error in $K$ caused by setting $e=0$ in
equation (\ref{eq:K}) is only 11\%.  We also assume that $M\ll
M_\ast$, so that equation (\ref{eq:K}) simplifies to
\be
K={m\over M_\ast}\left[2\pi GM_\ast\over P\right]^{1/3},
\label{eq:kdef}
\ee 
where $m\equiv M\sin i$ is the minimum companion mass, corresponding to an
orbit viewed edge-on.

Throughout this paper we shall assume that all of the stars in the
survey have mass equal to the Sun's, $M_\star=M_\odot$. This is not a
bad approximation since most radial-velocity surveys for low-mass
companions have focused on solar-type stars. In fact, our method is
easily generalized to the case where the survey stars have different
masses, but to do so we need to know the masses of {\em all} the stars
in the survey (not just the ones that have detected companions)---and
this extra complication did not seem worthwhile in this preliminary
analysis.\footnote{One of the largest relative errors caused by
setting $e=0$ and $M_\ast=M_\odot$ is for $\epsilon$ Eri
($M_\ast=0.8M_\odot$, $e=0.61$), where equation (\ref{eq:kdef}) with
$M_\ast=M_\odot$ yields a value for $K$ that is 46\% too small.}.

We restrict our attention to companions with minimum mass $m\le
m_\mathrm{max}\equiv 10\mj$, where $\mj$ is the Jupiter mass; this
cutoff hopefully minimizes the contamination of our sample by brown
dwarfs and is below the deuterium-burning threshold which sometimes is
taken to define the boundary between planets and stars. We also
restrict our attention to orbital periods $P>P_\mathrm{min}=2$ days,
corresponding to a semimajor axis of $6.7R_\odot=0.031$ AU; this limit
is small enough to include all the known planets.

We assume that the probability that a single star has a companion with
mass and orbital period in the range $[M,M+dM]$, $[P,P+dP]$ is given
by a power law,
\be
dp=C \left(M\over M_0\right)^{-\alpha} \left(P\over P_0\right)^{-\beta}
{dM\over M}{dP\over P},
\label{eq:lumfnnn}
\ee
where $C$, $\alpha$ and $\beta$ are constants to be determined, and
$M_0$ and $P_0$ are a fiducial mass and period, which we choose to be
$M_0=1.5 \mj$ and $P_0=\hbox{80 d}$ (the reasons for this choice are
outlined in the following subsection). If the distribution of
companion orbits is isotropic, the distribution of minimum mass
$m=M\sin i$ and period is given by
\be
dp=c \left(m\over M_0\right)^{-\alpha} \left(P\over P_0\right)^{-\beta}
{dm\over m}{dP\over P},
\label{eq:lumfn}
\ee 
where
\be
c=2^\alpha{\Gamma(1+\half\alpha)^2\over \Gamma(2+\alpha)}C,
\ee
$\Gamma(\cdot)$ is the gamma function, and $\alpha>-2$.

Initially we assume that the survey detects a companion if and only if
(i) the velocity amplitude $K$ exceeds a survey-dependent
detectability limit $K_D$; and (ii) its orbital period is shorter than
a survey-dependent upper limit $\pmax$. We expect that $\pmax$ will be
proportional to the duration of the survey, since typically at least
two orbits are required for a reliable detection. More realistic
smooth cutoffs to the detection efficiency are discussed in
\S\ref{sec:smooth}. Although the detection limits $K_D$ and $\pmax$
can be estimated from descriptions of the survey, we adopt the more
objective approach of determining them directly from the
maximum-likelihood analysis.

Let $x_i=\ln (m_i/M_0)$ and $y_i=\ln (P_i/P_0)$, $i=1,\ldots,N$, where
$m_i$ and $P_i$ are the minimum mass and period of the companions
detected in the survey.  Then the velocity amplitude $K_i$
(eq.~\ref{eq:kdef}) exceeds the detection limit $K_D$ if
\be x_i-\ffrac{1}{3}y_i > v\equiv
\ln\left(K_D\over28.4\,\hbox{m
s}^{-1}\right)-\ln\left(M_0\over\mj\right)+\ffrac{1}{3}\ln\left(P_0\over
1\hbox{ yr}\right).
\label{eq:vdef}
\ee
Similarly, the orbital period is less than the maximum detectable
period if
\be
y_i  <  u\equiv \ln\left(\pmax\over P_0\right).
\label{eq:udef}
\ee
The other constraints are 
\begin{eqnarray}
x_i&<&\xmax\equiv \ln(m_\mathrm{max}/M_0)=\ln
10+\ln(\mj/M_0),\nonumber \\
y_i&>&\ymin\equiv \ln(P_\mathrm{min}/P_0)=-5.207+\ln(1\hbox
{ yr}/P_0). 
\end{eqnarray}
The constants $\xmax$ and $\ymin$ are fixed, while the variables $u$
and $v$ are to be determined by the maximum-likelihood analysis.

{}From equation (\ref{eq:lumfn}) the expected number of companions in
the interval $dx\,dy$ in a survey of $N^\ast$ stars is
\be
\label{eqn:expnum}
n(x,y)dx\,dy = N^\ast p(x,y)dx\,dy\qquad\hbox{where}\qquad
p(x,y)=ce^{-\alpha x-\beta y}.
\ee
The likelihood function $L$ is the product of (i) the probability of
detecting $N$ companions with minimum masses $x_i$ and periods $y_i$;
and (ii) the probability of observing none elsewhere in the domain $D$
of $(x,y)$ space in which companions are detectable. Thus
\be
\label{eqn:like}
L=\prod_{i=1}^N n(x_i,y_i) \exp \left[ -\int_D dx\,dy\,n(x,y)\right],
\qquad \hbox{if all } (x_i,y_i)\in D,
\ee
and zero otherwise. The domain $D$ is $v+{1\over3}y<x<\xmax$,
$\ymin<y<\tilde{u}$, with
\be
\tilde{u}(u,v)\equiv\min \left[u,3(\xmax-v)\right].
\ee
We now substitute equation (\ref{eqn:expnum}) into equation
(\ref{eqn:like}) and take the log of the result,
\be
\label{eq:indsur}
\ln L = N \, \ln (cN^\ast)-\alpha\sum_{i=1}^N x_i-\beta
\sum_{i=1}^N y_i-cN^\ast f(\alpha,\beta,u,v).
\ee
Here
\be
\label{eqn:f}
f(\alpha,\beta,u,v) = \int_{-\infty}^{\tilde{u}} dy
\int_{v+y/3}^\infty dx\,g(\alpha,\beta,x,y), \ee
where
\be
\label{eq:gdef}
g(\alpha,\beta,x,y)=\left\{ \begin{array}{ll} e^{-\alpha x-\beta y} & 
\mbox{if $x<\xmax$ and $y>\ymin$,} \\ 0 
& \mbox{otherwise.} \end{array} \right.  
\ee
The integral yields 
\be 
f(\alpha,\beta,u,v)=\frac{3 e^{-\alpha
v}}{\alpha(\alpha+3\beta)} \left[ e^{-\frac{1}{3}\a3b\ymin} -
e^{-\frac{1}{3}\a3b\tilde{u}(u,v)}\right] +
\frac{e^{-\alpha\xmax}}{\alpha\beta}\left[e^{-\beta
\tilde{u}(u,v)}-e^{-\beta\ymin}\right], 
\ee 
if $v<\xmax-\ffrac{1}{3}\ymin$ and $u>\ymin$, and zero otherwise.

The best estimates for the fitted variables $c$, $\alpha$, $\beta$,
$u$ and $v$ correspond to the global maximum of $\ln L$. First, the
constant $c$ can be evaluated from
\be
\frac{\partial \ln L}{\partial c} = \frac{N}{c} - N^\ast f = 0 \;\;\;
\Longrightarrow \;\;\; c = \frac{N}{N^\ast f}.
\ee
Substituting this result into equation (\ref{eq:indsur}) yields
\be
\label{eq:indsurbis}
\ln L = N \left[ \ln \left( \frac{N}{f} \right) - 1 \right] -\alpha
\sum_{i=1}^N x_i - \beta \sum_{i=1}^N y_i.
\ee
To determine the best estimates for $u$ and $v$ we note that $\ln L$
depends on these parameters only through $f(\alpha,\beta,u,v)$, and
that $\ln L$ is maximized when $f$ is minimized. According to equation
(\ref{eqn:f}), $f$ depends on $u$ and $v$ only through the limits of
integration, that is, only through the shape of the domain $D$. Since
the integrand is non-negative, we minimize $f$ by making $D$ as small
as possible, so long as it still contains all the data points
$(x_i,y_i)$. This can be done by setting $v$ equal to the smallest
value of $x_i-{1 \over 3}y_i$ in the sample, and $u$ equal to the
largest value of $y_i$ in the sample.

The best estimates for the remaining parameters are given by
\begin{eqnarray}
\label{eqn:dalpha}
0=\frac{\partial \ln L}{\partial \alpha } &=& - \frac{N}{f} 
\frac{\partial f}{\partial \alpha} - \sum_{i=1}^{N} x_i, \\
\label{eqn:dbeta}
0=\frac{\partial \ln L}{\partial \beta } &=& - \frac{N}{f} 
\frac{\partial f}{\partial \beta} - \sum_{i=1}^{N} y_i,
\end{eqnarray}
which can easily be solved numerically. Table~\ref{tab:indsur} lists
the best estimates of the parameters for each survey. The value of the
normalization parameters $C$ is based on the fiducial mass and period
$M_0=1.5 \mj$ and $P_0=\hbox{80 d}$, which are chosen for reasons
outlined in the following subsection.

\begin{figure}
\begin{center}
\centerline{\epsfig{figure=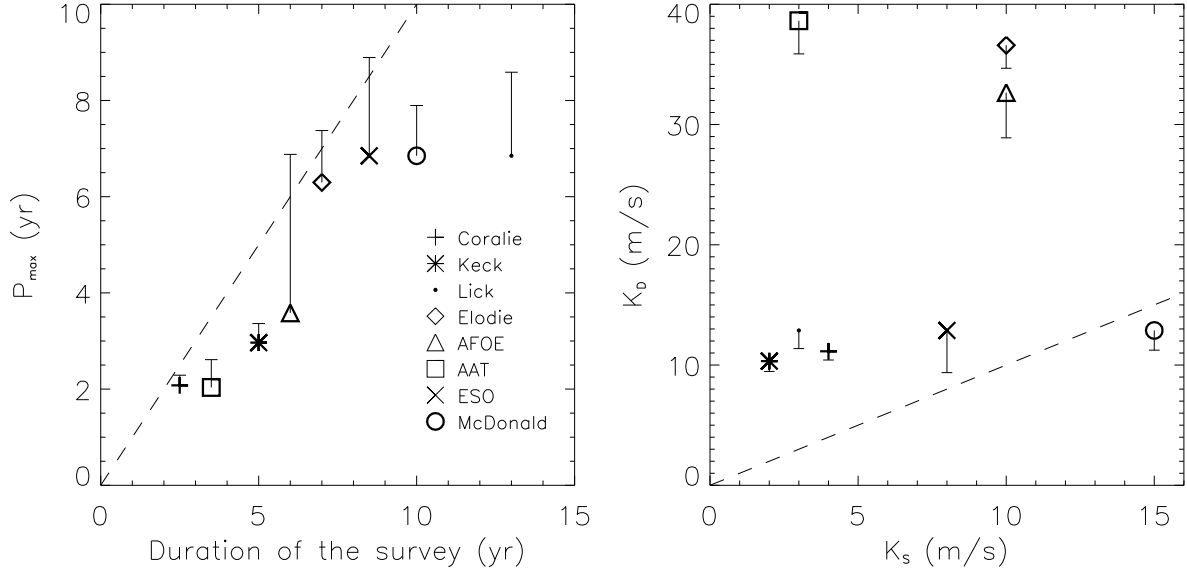}}
\caption{The left panel shows the correlation between the duration of
each survey (from Table \ref{tab:surveys}) and the longest period in
the sample. Note that Lick, ESO and McDonald programs have all
detected the same long-period planet: $\epsilon$ Eridani, which has an
inferred orbital period of $P=2502.1$ days. All of the points lie
below the dashed line ($x=y$), indicating that a reliable detection
requires following the star for more than one orbital period. The
right panel shows the correlation between the stated velocity
precision of each survey (from Table \ref{tab:surveys}) and the
smallest velocity amplitude of any of their detected planets. Almost
all of the points lie well above the dashed line, indicating that
determining a reliable orbit requires a velocity amplitude that is
significantly larger than the stated velocity precision. The exception
is the derived limit for McDonald; in this case $K_D$ is set by their
detection of a planet in the $\epsilon$ Eri system, for which our
approximations of a circular orbit and solar-mass star yield an
estimate for the velocity amplitude that is 46\% too low (see
footnote 1).}
\label{fig:pannel}
\end{center}
\end{figure}

Figure \ref{fig:pannel} shows the correlation between the duration of
each survey and the fitted value of $\pmax$ for that survey, as well
as the stated velocity precision $K_S$ for each survey and the fitted
detection limit $K_D$ for that survey (values taken from Tables
\ref{tab:surveys} and \ref{tab:indsur}). The detection limit is
generally a factor of three or so higher than the stated precision,
presumably because determining a reliable orbit is more difficult than
simply detecting the presence of a companion. For most of the surveys
there is an good correlation between duration and $\pmax$, with the
slope of the correlation indicating that approximately two orbital
periods of data are needed for a reliable detection.

\subsection{Generalization to multiple surveys}

\label{sec:mult}
\noindent
It is straightforward to expand the analysis of the previous Section
to multiple surveys, which we label by $j=1,\ldots,J$. The three
parameters describing the companion distribution, $\alpha$, $\beta$
and $c$, are now derived from the entire sample of known companions
from all surveys, while the parameters $u_j$ and $v_j$ that describe
the period and radial-velocity thresholds are different for each
survey. Thus the expected number of companions to be discovered in the
interval $dx\,dy$ in survey $j$ is
\be
\label{eqn:expnuma}
n_j(x,y)dx\,dy = N^\ast_j p(x,y)dx\,dy,
\ee
where $p(x,y)$ is defined in equation (\ref{eqn:expnum}) and
$N^\ast_j$ is the number of stars in survey $j$. Equation
(\ref{eqn:like}) becomes
\be
\label{eqn:likej}
L_j=\prod_{i=1}^{N_j} n_j(x_{i,j},y_{i,j}) \exp \left[ -\int_{D_j}
n_j(x,y)dxdy \right], 
\ee
and the likelihood is given by 
\be
\label{eqn:Ltot}
\ln L = \sum_{j=1}^{J} \ln L_j.
\ee
Again, the integration constant can be eliminated from
equation (\ref{eqn:Ltot}) 
\be
\frac{\partial \ln L}{\partial c} = \sum_{j=1}^J
\left( \frac{N_j}{c} - N_j^\ast f_j \right) = 0 \;\;\; \Longrightarrow
\;\;\; c = \frac{\sum_{j=1}^J N_j}{\sum_{j=1}^J N_j^\ast f_j },
\ee
and the likelihood becomes
\be
\ln L = \sum_{j=1}^{J} N_j \left[ \ln N_j^\ast
+\ln\sum_{j=1}^J N_j-\ln \sum_{j=1}^J N_j^\ast f_j -1\right]
-\alpha\sum_{j=1}^J\sum_{i=1}^{N_j}
x_{i,j} - \beta\sum_{j=1}^J\sum_{i=1}^{N_j} y_{i,j}.
\ee
As before, $v_j$ is set equal to the smallest value of
$x_{i,j}-{1\over 3}y_{i,j}$ in survey $j$, and $u_j$ is set equal to
the largest value of $y_{i,j}$ in survey $j$.

The surveys listed in Table \ref{tab:surveys} have discovered 90
companions, although several appear in more than one survey so there
are only 63 distinct companions. Companions discovered in multiple
surveys are counted in each survey where they appear; this approach
leads us to underestimate the statistical uncertainties in our
parameters somewhat (probably by about a factor of
$(90/63)^{1/2}=1.2$), but avoids the systematic bias that would be
created by counting the companions only once and discarding them from
the other surveys. A conservative alternative approach is to use only
the Coralie survey for parameter estimation (top lines of Tables
\ref{tab:surveys} and \ref{tab:indsur}).

The values of the normalization parameters $c$ and $C$ quoted in this
paper are based on the fiducial mass and period $M_0=1.5 \mj$ and
$P_0= 80\hbox{ d}$. These values are chosen to minimize the
uncertainty in $\ln c$. If the uncertainties are small, this
requirement is equivalent to choosing $M_0$ and $P_0$ so that the
covariances between $c$ and the exponents $\alpha$ and $\beta$ vanish.

\begin{figure}
\begin{center}
\centerline{\epsfig{figure=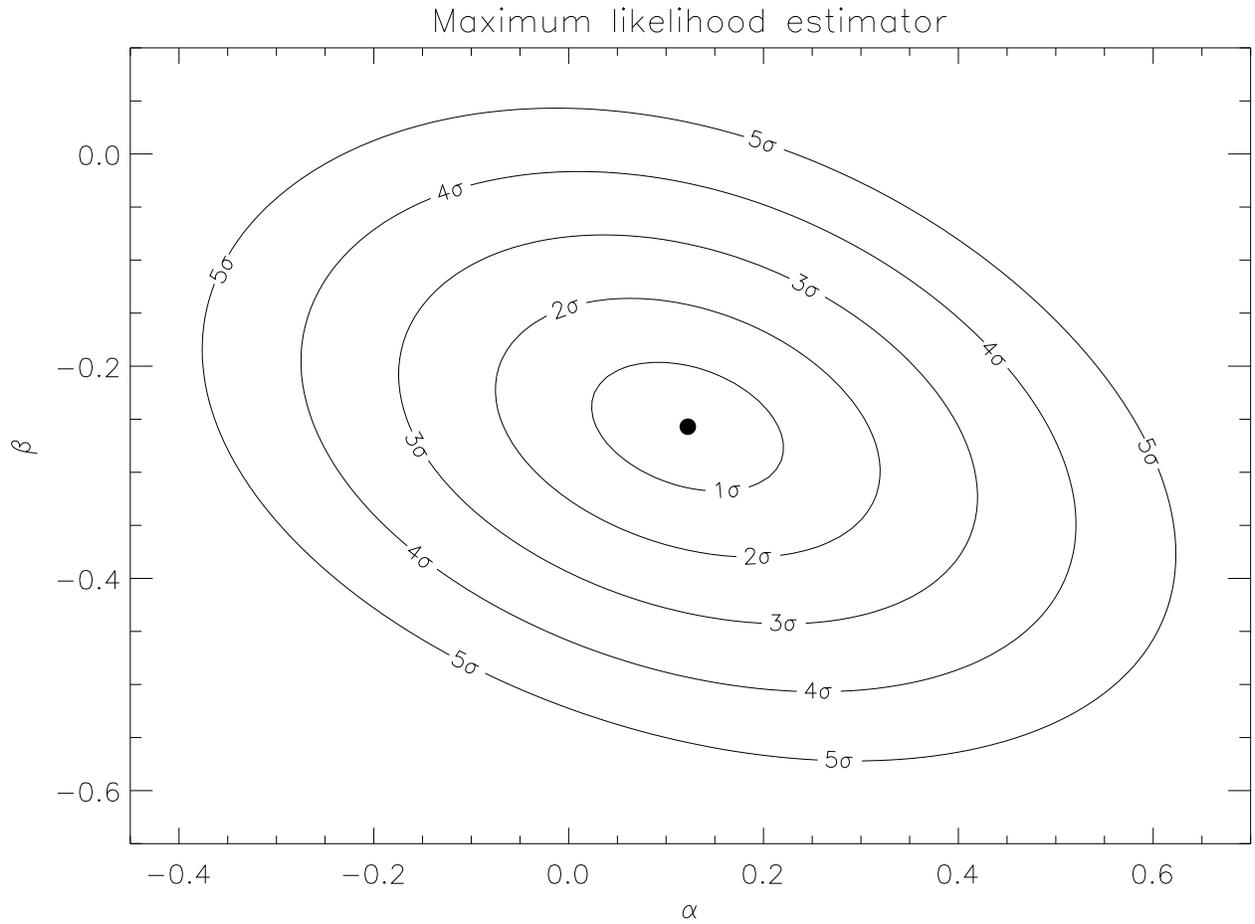}}
\caption{Contours of constant likelihood for the combination of all
eight surveys. The maximum of $L$ is located at the filled circle,
$\alpha=0.12\pm0.10$, $\beta=-0.26\pm0.06$. The contour levels
represent ``$n$--$\sigma$'' confidence regions, in which the
likelihood function is smaller than its maximum value by
$\exp(-n^2/2)$.}
\label{fig:contab}
\end{center}
\end{figure}

\begin{figure}
\begin{center}
\centerline{\epsfig{figure=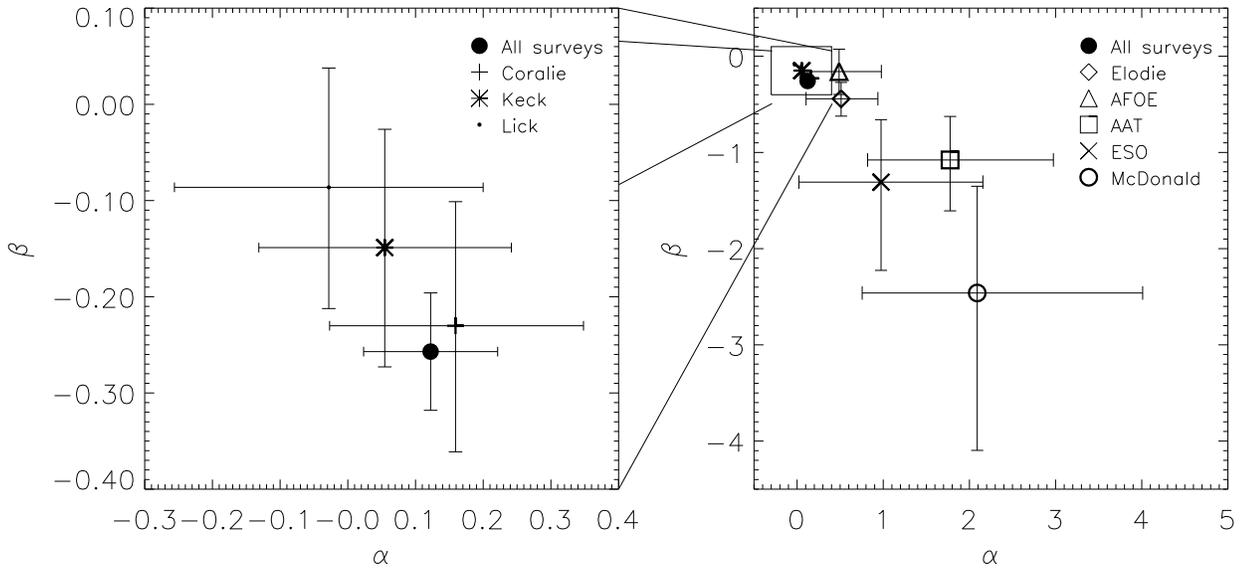}}
\caption{The estimators of the exponents of the mass and period
distribution, $\alpha$ and $\beta$, for the combined eight surveys
(filled circles) compared with their values for each individual
survey. The estimates are approximately consistent given the
uncertainties.}
\label{fig:alphabeta}
\end{center}
\end{figure}

The likelihood analysis yields 
\begin{eqnarray}
\label{eqn:alpha}
\alpha &=& 0.12 \pm 0.10, \\
\label{eqn:beta}
\beta &=& -0.26 \pm 0.06, \\
\label{eqn:c}
c &=& 1.7^{+0.19}_{-0.17} \times 10^{-3} \\
\label{eqn:cc}
C &=& 1.8^{+0.19}_{-0.18} \times 10^{-3} 
\end{eqnarray}
where the confidence limits correspond to $\ln L=(\ln L)_\mathrm{max}
- 0.5$. With these estimators at hand, it is straightforward to plot
the likelihood as a function of the exponents $\alpha$ and $\beta$
(Figs.~\ref{fig:contab} and \ref{fig:alphabeta}). Figure
\ref{fig:contab} indicates that there is a significant covariance
between the exponents $\alpha$ and $\beta$ that characterize the mass
and period distributions (correlation coefficient $r=-0.32$). This
correlation arises because of the selection effects on velocity
amplitude, and demonstrates that both distributions should be fitted
simultaneously.

Table~\ref{tab:pbd} shows the difference between the number of
predicted companions and the number of actual detections; these are
generally in good agreement except for the McDonald survey, which is
discussed further in \S\ref{sec:extrap}.

This table also lists the number of predicted brown dwarfs ($10\mj< M
< 80\mj$) in each sample, assuming that the planetary mass function
extends to $80\mj$, and the number of brown dwarfs actually
discovered. As many authors have pointed out, the small number of
brown dwarf discoveries strongly suggests that the mass function we
have derived cannot be extrapolated to brown dwarf masses.

\subsection{Generalization to a smooth cutoff \label{sec:smooth}}
\noindent
A sharp cutoff in the detectability of planets at radial velocity
$K_D$ and period $\pmax$ is not very realistic. A better approximation
is to replace the sharp cutoffs in equations (\ref{eq:indsur}) and
(\ref{eqn:f}) with smooth functions. We can do this by replacing
$f(\alpha,\beta,u,v)$ with
\begin{eqnarray}
\label{eq:smooth}
f_s(\alpha,\beta,u,v)& = &\int_{\ymin}^\infty dy\,h_u(u-y) 
\int_{-\infty}^{\xmax}dx\,h_v(x-\ffrac{1}{3}y-v)e^{-\alpha x-\beta y},
\nonumber \\ 
& = & \int_{-\infty}^\infty dy\, h_u(u-y)
\int_{-\infty}^\infty dx\,h_v(x-\ffrac{1}{3}y-v)g(x,y),
\end{eqnarray}
where $g(x,y)$ is defined by equation (\ref{eq:gdef}). 

The functions $h_u(\cdot)$ and $h_v(\cdot)$ are measures of the
detection efficiency of the survey as a function of orbital period and
velocity amplitude. The function $h_u(s)$ approaches 0 as $s\to
-\infty$ and $1$ as $s\to\infty$; we shall assume that $h_u(s)-\half$
is an odd function of $s$ so that $h_u(0)=\half$, with similar
assumptions for $h_v$. In the limit where $h_u$ and $h_v$ are step
functions we recover equation (\ref{eqn:f}). Thus $v$ and $u$ are
still defined by equations (\ref{eq:vdef}) and (\ref{eq:udef}), except
that $K_D$ and $\pmax$ are interpreted as the velocity amplitude and
orbital period at which the detection efficiency falls to 50\%.

Let 
\be
h_u(s) = \int_{-\infty}^s b_u(s')ds', \qquad h_v(t) = \int_{-\infty}^{t}
b_v(t')dt';
\ee
then equation (\ref{eq:smooth}) can be rewritten as
\begin{eqnarray}
\label{eqn:fs2}
f_s(\alpha,\beta,u,v) &=& \int_{-\infty}^{\infty}
ds'\,b_u(s')\int_{-\infty}^{\infty} 
dt'\,b_v(t')\int_{-\infty}^{u-s'}dy \int_{t'+y/3+v}^{\infty}
dx\,g(x,y)\nonumber \\
&=& \int_{-\infty}^{\infty} ds\,b_u(s)\int_{-\infty}^{\infty}
dt\,b_v(t)f(\alpha,\beta,u-s,v+t),
\end{eqnarray}
where the second line follows from equation (\ref{eqn:f}), and we have
dropped the primes on the dummy variables. These integrals are easy to
evaluate numerically.

\begin{figure}
\begin{center}
\centerline{\epsfig{figure=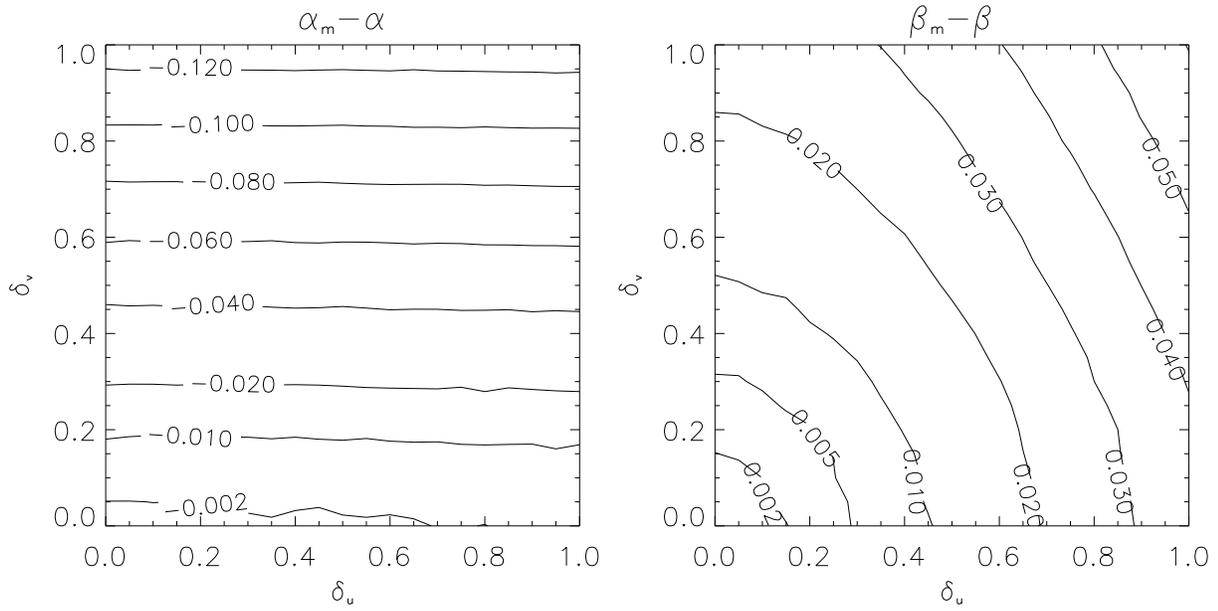}}
\caption{The difference of the modified best parameters $\alpha_m$ and
$\beta_m$ (which depend on the threshold widths $\delta_u$ and
$\delta_v$) and the nominal values of $\alpha$ and $\beta$ as quoted
in eqs.~(\ref{eqn:alpha}) and (\ref{eqn:beta}).}
\label{fig:alpha}
\end{center}
\end{figure}

We shall choose 
\be 
b_u(s)={1\over\sqrt{2\pi}\delta_u}\exp\left(-{s^2\over
2\delta_u^2}\right), \qquad h_u(s)=\half + \half\erf\left(s\over
\sqrt{2}\delta_u\right), 
\ee 
with a similar choice for $h_v(t)$. We call $\delta_u$ and $\delta_v$
the threshold widths. Figure~\ref{fig:alpha} shows the effect of
non-zero threshold widths on the slope estimators $\alpha$ and
$\beta$. In the arbitrary but plausible case where the detection
efficiencies $h$ drop from ${3\over4}$ to ${1\over4}$ over a factor of
two in period or velocity amplitude, we have $\delta_u$,
$\delta_v=0.51$. In this case the best-fit value of $\alpha$ is
shifted downward by 0.04 and the best fit for $\beta$ is shifted
upwards by about 0.020. These changes are significant but relatively
modest; since the appropriate values of the threshold widths are
difficult to estimate, we shall not attempt to correct for this
effect.

\section{Discussion}
\label{sec:disc}

\subsection{Comparison to other estimates of the mass distribution}

\noindent
We have found that the distribution of companion masses below $10\mj$
is approximately flat in $\log M$, or slightly rising towards small
masses, i.e., the exponent $\alpha$ is small and positive
($\alpha=0.12\pm0.1$, eq.~\ref{eqn:alpha}). The distribution of
companion masses has already been examined by a number of authors,
most of whom have reached similar conclusions
(\cite{maz98,mar98,maz99,sb00,sb01,jor01,zuc01}). Our approach offers
several advantages over the variety of methods used in these papers:
(i) we correct for selection effects in period and velocity amplitude,
(ii) we account for the coupling between the orbital period
distribution and mass distribution, (iii) we estimate the sensitivity
in radial velocity or maximum period (our parameters $K_D$ and
$\pmax$) self-consistently from the data, and (iv) we determine the
normalization of the mass distribution, not just its shape.

\subsection{Extrapolations}
\label{sec:extrap}

\noindent
It is interesting to investigate the implications of extrapolating the
mass and period distributions that we have derived. If we assume that
our maximum-likelihood distribution
(eqs. \ref{eqn:alpha}--\ref{eqn:cc}) applies in the mass range
$10\mj<m<80\mj$ that is usually associated with brown dwarfs, we
predict that the Coralie and Keck surveys should have discovered,
respectively, 13 and 8 companions in this range (Table \ref{tab:pbd});
in fact these two surveys found only one companion each. This result
confirms the finding of several authors
(\cite{bas97,may98,maz98,maz99,jor01}) that there is a cutoff in the
power-law distribution of companion masses at $m\gtrsim 10\mj$, and a
``brown-dwarf desert'' between $\sim 10\mj$ and $\sim 100\mj$ in which
few companions exist at semi-major axes less than a few AU.

The average number of planets per star with masses between $M_1$ and
$M_2$ and periods between $P_1$ and $P_2$ is given by equation
(\ref{eq:lumfnnn}):
\be
N={C\over\alpha\beta}\left[\left(M_0\over M_1\right)^\alpha-
\left(M_0\over M_2\right)^\alpha\right]\left[\left(P_0\over
P_1\right)^\beta-\left(P_0\over P_2\right)^\beta\right]. 
\label{eq:npm} 
\ee
Thus, for example, in our best-fit model
(eqs. \ref{eqn:alpha}--\ref{eqn:cc}), the expected number of planets
per star with periods between 2 days and $10\hbox{ yr}$ and masses
between $M$ and $10\mj$ is
\be
N=0.129\left[\left(M_0/ M\right)^{0.12}-0.755\right].
\ee
For $M=\mj$, $N=0.033$; thus about three percent of solar-type stars
have a planet of Jupiter mass or larger in this period range. If we
make the large extrapolation to Earth-mass planets ($M=0.003\mj$) we
find $N=0.172$; in this case, more than 15\% of stars would have an
Earth-mass or larger companion.

If we extrapolate to larger orbital periods, we find that the number
of companions in a given mass range with periods between 2 days and 5
yr would be about 0.28 times the number with periods between 5 yr and
1000 yr; in this case a significant fraction of Jupiter-mass planets
would have orbital periods short enough to be detected in existing
radial-velocity surveys.

\subsection{Comparison with the solar nebula}

\noindent
The mass and period distribution (\ref{eq:lumfnnn}) can be used to
determine the total surface density in planets less massive than
$M_{\rm max}$, assuming that the central star has mass $1M_\odot$:
\be
\Sigma(a)={3C\over 4\pi(1-\alpha)}{M_0\over (1\hbox{ AU})^2}
\left(M_{\rm max}\over M_0\right)^{1-\alpha}\left(P_0\over1
\hbox{ yr}\right)^{\beta} \left(1\hbox{ AU}\over a\right)^{2+3\beta/2}. 
\ee
For our best-fit model,
\be
\Sigma(a)=50\hbox{ g cm}^{-2} \left(M_{\rm max}\over 
10 \mj\right)^{0.9}\left(1\hbox{ AU}\over a\right)^{1.6};
\label{eq:sigobs}
\ee
this can be compared to the gas and dust densities required in the
minimum solar nebula (e.g. \cite{hay81})
\begin{eqnarray}
\Sigma_{\rm gas}(a) & = & 1.7\times10^3\hbox{ g cm}^{-2}
\left(1\hbox{ AU}\over a\right)^{1.5};\nonumber \\
\Sigma_{\rm dust}(a) & = & 7.1\hbox{ g cm}^{-2}
\left(1\hbox{ AU}\over a\right)^{1.5}.
\label{eq:sigtheory}
\end{eqnarray}
The agreement of the exponents in equations (\ref{eq:sigobs}) and
(\ref{eq:sigtheory}) is striking and perhaps surprising, given that
many theorists believe that the extrasolar giant planets must have
formed at much larger radii and migrated to their present locations,
while the planets in our solar system have suffered little or no
migration.

\subsection{Summary} 

\begin{figure}
\begin{center}
\centerline{\epsfig{figure=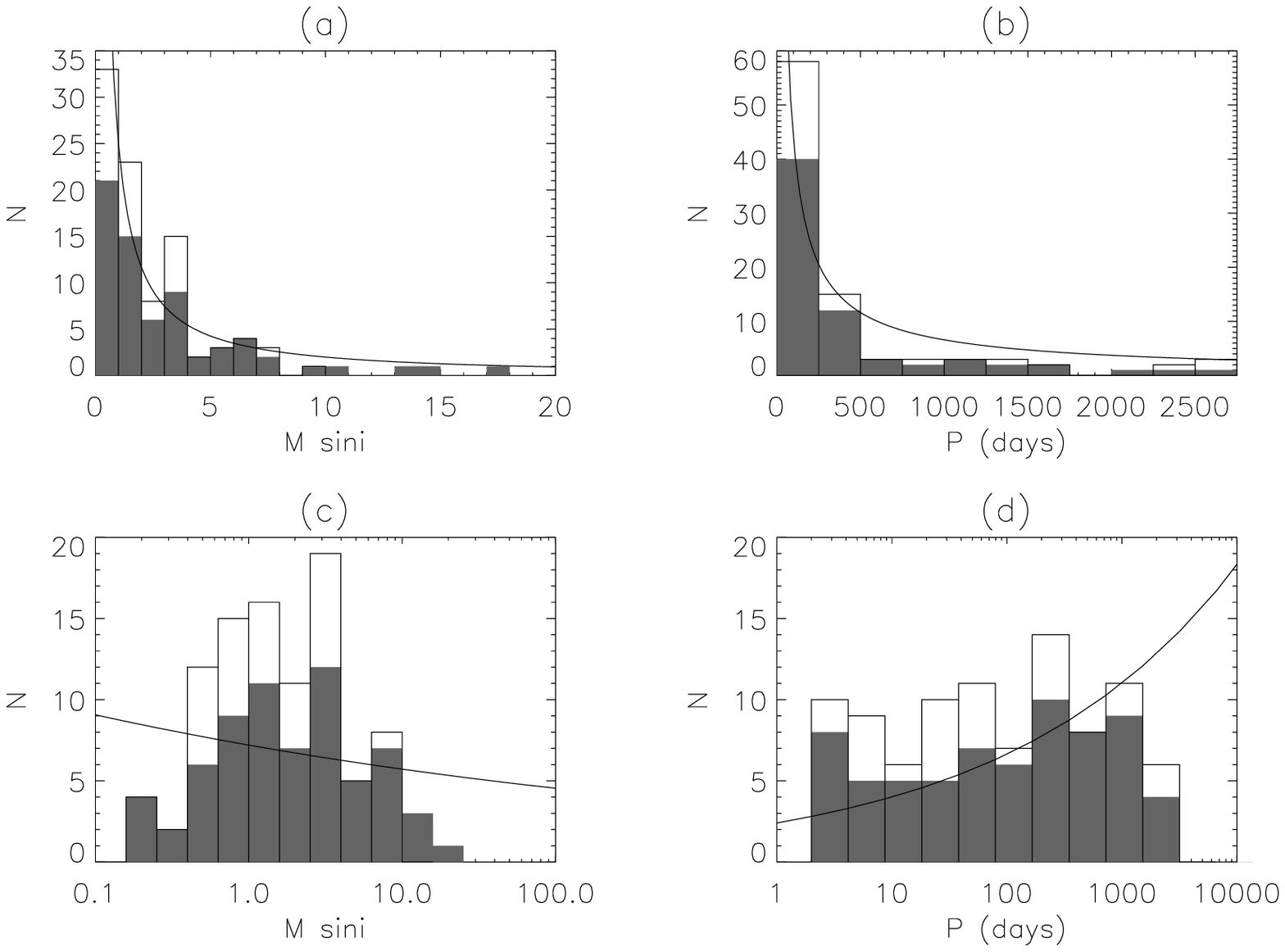}}
\caption{The minimum mass (left panels) and period (right panels)
distributions, on linear (top) and logarithmic (bottom) scales. The
gray histograms represent all objects, including those with $M\sin
i>10\mj$, found in the radial-velocity surveys in Table
\ref{tab:surveys} (67 objects). The unshaded histograms include
duplicate discoveries of the same object in different surveys, as
discussed in \S \ref{sec:mult} (94 objects). The curves show the
predictions of our best-fit model, given by equations
(\ref{eqn:alpha})--\ref{eqn:cc}) and (\ref{eq:npm}). As discussed in
the paper, the histograms are subject to selection effects at small
mass and large period, and there is evidence for a cutoff to the mass
distribution above $\sim 10\mj$. }
\label{fig:bds}
\end{center}
\end{figure}

Figure~\ref{fig:bds} shows the minimum-mass and period distributions
of all of the substellar companions found in the surveys in Table
\ref{tab:surveys}, along with the simple power-law models that we have
used to fit these distributions. The effects of the selection effects
at small mass and large period, and the evidence for a cutoff above
$\sim 10\mj$, are evident in the bottom panels.

We have described a simple maximum-likelihood method that determines
the mass and period distributions of extrasolar planets discovered in
multiple surveys. Our method determines and accounts for selection
effects on velocity amplitude and period from the data themselves,
without relying on the nominal survey parameters. Our best-fit model
is defined by equations (\ref{eq:lumfnnn}) and equations
(\ref{eqn:alpha})--(\ref{eqn:cc}).

We are indebted to W. Cochran, G. Marcy and M. Mayor for communicating
some of the details of their surveys that were used in this
statistical analysis. We are particularly grateful to David Weinberg
for pointing out a normalization error in an early version of this
paper. This research was supported in part by NASA grant NAG5-10456,
and by a European Space Agency fellowship.

{}
\begin{table}
\begin{center}
\begin{tabular}{lccccl}
\hline
\hline
Survey   & $K_\mathrm{S}$ & duration & number of & number of dis- &
references \\
         & (m/s)  & (yr) & stars observed & covered planets & \\
\hline
Coralie  &    4   &  2.5 & 1000		& 23 & \cite{udr01} \\
Keck     &  2--5  &    5 &  530 	& 22 & \cite{vog00} \\
Lick     & 3--10  &   13 &  300		& 17 & \cite{cum99} \\
Elodie   &   10   &    7 &  320 	& 13 & \cite{udr01} \\
AFOE     &   10   &    6 &  100 	&  7 & \cite{nis99} \\
AAT	 &    3   &  3.5 &  200 	&  4 & \cite{tin01} \\
ESO      & 8--15  &  8.5 &   40 	&  2 & \cite{end00} \\
McDonald & 15--20 &   10 &   73 	&  2 & \cite{coc00} \\
\hline
\hline
\end{tabular}
\caption{Characteristics of eight radial-velocity surveys of
extrasolar planets: stated velocity precision of the survey
($K_\mathrm{S}$), duration of the program, number of stars observed to
date, number of planetary companions discovered ($M\sin i<10\mj$), and
references for each survey.}
\label{tab:surveys}
\end{center}
\end{table}
\begin{table}
\begin{center}
\begin{tabular}{lccccc}
\hline
\hline
Survey & $\alpha$ & $\beta$ & $C\times 10^4$ & $\pmax$ &
$K_\mathrm{D}$ \\  
&&&& (yr) & ($\mathrm{m\ s}^{-1}$) \\
\hline
Coralie  & $0.16 \pm 0.19$	    & $-0.23 \pm 0.13$ \
	 & $11.5^{+2.7}_{-2.3}$   & $2.1+0.2$ & $11.1-0.7$\\
Keck     & $0.06 \pm 0.19$ 	    & $-0.15 \pm 0.12$ \
	 & $18.1^{+4.2}_{-3.7}$   & $3.0+0.4$ & $10.3-0.9$\\
Lick     & $-0.03 \pm 0.23$	    & $-0.09 \pm 0.12$ \
	 & $22.2^{+5.9}_{-5.0}$  & $6.9+1.7$ & $12.9-1.5$\\
Elodie   & $0.51^{+0.43}_{-0.41}$   & $-0.44^{+0.17}_{-0.18} $ \
	 & $32.5^{+11.5}_{-9.5}$   & $6.3+1.1$ & $36.6-1.9$\\
AFOE     & $0.49^{+0.49}_{-0.46} $  & $-0.16^{+0.23}_{-0.24} $ \
	 & $55.9^{+26.7}_{-20.4}$   & $3.6+3.3$ & $32.7-3.8$\\
AAT      & $1.78^{+1.2}_{-0.96} $   & $-1.08^{+0.45}_{-0.53}$ \
	 & $31.4^{+19.8}_{-14.1}$  & $2.0+0.6$ & $38.6-2.8$\\
ESO      & $0.98^{+1.18}_{-0.95}$   & $-1.31^{+0.65}_{-0.92}$ \
	 & $4.2^{+15.8}_{-3.9}$  & $6.9+2.0$ & $12.9-3.5$ \\
McDonald & $2.09^{+1.92}_{-1.34}$   & $-2.46^{+1.11}_{-1.64}$ \
	 & $0.1^{+2.5}_{-0.09} $   & $6.9+1.0$ & $12.9-1.6$\\
\hline
\end{tabular}
\caption{For each survey, best estimates for the exponents $\alpha$
and $\beta$ and the normalizing constant $C$ in equation
(\ref{eq:lumfnnn}), the maximum detectable period $\pmax$ and the
velocity precision $K_\mathrm{D}$ in $\mathrm{m\ s}^{-1}$. The
fiducial mass and period are $M_0=1.5\mj$, $P_0=80$ d. The errors on
$\pmax$ and $K_D$ are one-sided, since the maximum likelihood is
achieved when these parameters equal the largest period and smallest
velocity amplitude found in the survey.}
\label{tab:indsur}
\end{center}
\end{table}
\begin{table}
\begin{center}
\begin{tabular}{lcccc}
\hline
\hline
Survey   & number of pre- & number of dis-  & number of predic- & number of
discovered  \\ 
         & dicted planets & covered planets & ted brown dwarfs & brown dwarfs
\\
\hline
Coralie  & $35^{+6}_{-4}$  & 23 & $13^{+7}_{-4}$ & 1 \\
Keck     & $21^{+3}_{-2}$  & 22 & $8^{+4}_{-3}$  & 1 \\
Lick     & $13 \pm 2$      & 17 & $6^{+4}_{-2}$  & 1 \\
Elodie   & $9^{+2}_{-1}$   & 13 & $6^{+4}_{-2}$  & -- \\
AFOE     & $3 \pm 0$       &  7 & $2 \pm 1$      & -- \\
AAT      & $4 \pm 1$	   &  4 & $3 \pm 1$      & -- \\
ESO      & $2 \pm 0$	   &  2 & $1 \pm 0$   	 & -- \\
McDonald & $3^{+1}_{-0}$   &  2 & $1 \pm 1$ 	 & -- \\
\hline
\hline
\end{tabular}
\caption{Comparison of predicted and observed numbers of
companions.}
\label{tab:pbd}
\end{center}
\end{table}
\clearpage

\begin{thebibliography}{}
%
\bibitem[Basri \& Marcy 1997]{bas97} Basri, G., \& Marcy, G. W. 1997,
in Star Formation Near and Far, AIP Conf. Proc. 393, eds. S.  Holt \&
L. G. Mundy (New York, AIP), 228
%
\bibitem[Cochran \etal 2000]{coc00} Cochran, W. D., Hatzes, A.P., \&
Paulson, D.B. 2000, in Planetary Systems in the Universe: Observation,
Formation and Evolution, IAU Symp. 202, eds. A. Penny, P. Artymowicz,
A.-M. Lagrange and S. Russell (ASP Conf. Ser.) in press
%
\bibitem[Cumming \etal 1999]{cum99} Cumming, A., Marcy, G. W., \&
Butler, R. P. 1999, \apj, 526, 890
%
\bibitem[Endl \etal 2000]{end00} Endl, M., K\"{u}rster, M., \& Els,
S. 2000, \aap, 362, 585
%
\bibitem[Hayashi 1981]{hay81} Hayashi, C. 1981,
Progr. Theor. Phys. Suppl., 70, 35
%
\bibitem[Jorissen \etal 2001]{jor01} Jorissen, A., Mayor, M., \& Udry,
S. 2001, submitted to \aap, astro-ph/0105301
%
\bibitem[Marcy \& Butler 1998]{mar98} Marcy, G. W., \& Butler,
R. P. 1998, \araa, 36, 57
%
\bibitem[Mayor \etal 1998]{may98} Mayor, M., Queloz, D., \& Udry,
S. 1998, in Brown Dwarfs and Extrasolar Planets, ASP Conf. Ser. 134,
eds. R. Rebolo, E. L. Martin, \& M. R. Zapatero-Osorio (San Francisco:
ASP), 140
%
\bibitem[Mazeh 1999]{maz99} Mazeh, T. 1999, Physics Reports, 311, 317
%
\bibitem[Mazeh \etal 1998]{maz98} Mazeh, T., Goldberg, D., \& Latham,
D. W. 1998, \apj, 501, L199
%
\bibitem[Nisenson \etal 1999]{nis99} Nisenson, P., Contos, A.,
Korzennik, S., Noyes, R., \& Brown, T. 1999, in Precise Stellar Radial
Velocities, ASP Conf. Ser. 185: IAU Colloq. 170, eds. J. B. Hearnshaw
and C. D. Scarfe (San Francisco: ASP), 143
%
\bibitem[Stepinski \& Black 2000]{sb00} Stepinski, T. F., \& Black,
D. C. 2000, \aap, 356, 903
%
\bibitem[Stepinski \& Black 2001]{sb01} Stepinski, T. F., \& Black,
D. C. 2001, \aap, 371, 250
%
\bibitem[Tinney \etal 2001]{tin01} Tinney, C. G., Butler, R. P.,
Marcy, G. W., Jones, H., Penny, A., Vogt, S., Apps, K., Henry,
G. W. 2001, \apj, 551, 507
%
\bibitem[Udry \etal 2001]{udr01} Udry, S., Mayor, M., \& Queloz,
D. 2001, in Planetary Systems in the Universe: Observation, Formation
and Evolution, IAU Symp. 202, eds. A. Penny, P. Artymowicz,
A.-M. Lagrange and S. Russell (ASP Conf. Ser.) in press
%
\bibitem[Vogt \etal 2000]{vog00} Vogt, S., Marcy, G., \& Butler,
R. P. 2000, \apj, 536, 902
%
\bibitem[Zucker \& Mazeh 2001]{zuc01} Zucker, S., \& Mazeh, T. 2001,
submitted to \apj, astro-ph/0106042

\end{thebibliography}
\end{document}